\documentclass[12pt]{book}

\usepackage[dvips]{graphicx,color}
\usepackage{makeidx,tsukuba,amssymb}

\makeauthorindex
\makeindex

\begin{document}

\BookTitle{\itshape The 28th International Cosmic Ray Conference}
\CopyRight{\copyright 2003 by Universal Academy Press, Inc.}
\pagenumbering{arabic}

\chapter{ANTARES Status Report}

\author{%
%
%
Teresa Montaruli$^1$ for the ANTARES Collaboration$^2$\\
{\it (1) Universit\'a di Bari and INFN, Via Amendola, 173, I-70126, Bari} \\
{\it (2) http://antares.in2p3.fr}
}

\section*{Abstract}
The ANTARES Collaboration is building a neutrino telescope 
2400 m below the Mediterranean sea close to the Southern French coast. 
The site is already linked to the shore station by a 40 km-long 
electro-optical cable (EOC) which transmits power and data. 
A prototype line and an instrumentation line 
for monitoring environmental parameters have been successfully deployed 
and connected to the EOC via the junction box, using the IFREMER manned 
submarine.
The Collaboration, after
years of dedicated R\&D and deployments of prototype lines, is now
ready to deploy the detector starting in spring 2004.

\section{The ANTARES neutrino telescope}

The ANTARES (Astronomy with a Neutrino Telescope and Abyss environmental
RESearch) project started in 1996 and involves physicists and engineers 
from France, 
Germany, Italy, Russia, Spain, The Netherlands and the United Kingdom. 
ANTARES aims to detect neutrinos with $E_{\nu}\gtrsim 10$~GeV 
in order to investigate $\nu$ astrophysics, 
dark matter in the form of weakly interacting massive particles (WIMPs), 
monopoles and $\nu$ oscillations. Cherenkov light produced by
relativistic charged particles is detected by a 3D-array of
optical modules (OMs), 
pressure resistant glass spheres containing phototubes (PMTs).
Photon arrival times and PMT charge amplitudes
allow track and energy reconstruction. 

Current predictions and upper limits from previous generation and
currently running telescopes indicate that the expected signal from 
cosmic neutrino sources require very large detectors to be observed. 
About 200 $\nu$-induced upward-going muons/km$^{2}$/yr 
are expected for a diffuse flux from an isotropic 
distribution of optically-thin extra-galactic sources equal 
to the Waxman \& Bahcall limit of $4.5 \cdot 10^{-8} E^{-2}$~GeV$^{-1}$ 
cm$^{-2}$ s$^{-1}$ sr$^{-1}$~[9]. Galactic source luminosities
$\gtrsim 10^{35}$ erg/s, achievable in the presence of compact accelerators
and intense magnetic fields, are required to produce 
a rate larger than 10 events/yr from 100~TeV neutrinos in a km$^2$ array.
In 1 year of data-taking the ANTARES expected sensitivity will
surpass that of current arrays and that expected for 600 live-days
of AMANDA II~[2].
The success of this experiment, 
with an effective area $>0.02$~km$^2$ for $E_{\nu} > 10$~TeV and well 
reconstructed events,
will be a milestone demonstrating the feasibility of an underwater 
$km^3$ detector in the Mediterranean, complementing a similar array in 
the South Polar ice. Two $km^3$-size detectors, one in each hemisphere, 
are needed to cover the whole sky
(including the Galactic Centre which is not accessible from the South 
Pole using 
upward-going neutrinos), and guarantee a cross-check of systematic errors 
arising from different Cherenkov media properties.

The ANTARES site (42$^{\circ}$ 50'N, 6$^{\circ}$ 10'E) is 
well-shielded from the atmospheric muon background by 2400~m of sea
water and
has been selected after an intense program of sea campaigns dedicated to
measurements of water transmission properties. The absorption length
is about 60~m at 470~nm and
mainly determines the size of the instrumented region and the PMT spacing. 
The effective scattering length~\footnote{Effective means that the 
scattered photon distribution, forward peaked in sea water, 
is accounted for.} is more than 200~m, this is
considerably larger and less depth dependent
than that of ice. 
The optical background, which is absent in ice, is due to $\beta$ decays 
of $^{40}K$ and to a continuous bioluminescence rate which combine to give
a rate of about 60~kHz on a 10'' PMT.
Occasional, short bioluminescence bursts momentarily increase the
rate up to MHz. 
These bursts induce a dead-time of 5\% per PMT, however the detector
dead-time from this source is far less due to the requirement for
coincidences.
The average light transmission loss of an OM
due to bio-fouling and sedimentation is $<2\%$ at its equator 
after 1~year from deployment, this saturates with time~[1].

Twelve lines will be deployed each with 75 OMs mounted in 
25 triplets (storeys).
Each OM has a 10'' Hamamatsu PMT~[10] oriented 
at $45^{\circ}$ from the vertical. Storey separation is 14.5~m giving 
a total implemented height of about 350~m which starts 
100~m above the sea bed. Lines are kept taut by buoys and are at an average 
distance of $\sim 65$~m from one another in an octagonal configuration. 
The production of the 900 OMs started in spring 2002.
The readout of each PMT signal is shared by 2 Analog Ring Sampler ASICs
which provide the analog signal time and charge digitization. 
The ARS implements a waveform (WF) shape-sensitive discrimination 
to distinguish single photoelectron-like pulse shapes 
(more than 98\% of events) from larger pulses.  
Tests have shown that an overall time resolution of $\sim 1$ ns can be achieved
mainly limited by the transit time spread (TTS) of the PMTs~[4].
The ARS's together with a compass and tilt meters or hydrophones for the 
line positioning
are located inside a titanium container which is common to each storey. 
Considering a singles rate of ~70 kHz from $^{40}K$ and
bioluminescence rate of 2\% of WF events, the typical 
data rate to shore will be $\sim 7$ MB/s/PMT;
a $\sim 100$ PC farm on shore will process data. 
The DAQ system design is given in~[3].
Time calibrations are critical for the ANTARES experiment: a system of
LED and laser beacons and a clock calibration system~[6] 
will allow a relative time precision
between OMs of $\sim 0.5$ ns 
and the absolute time will be determined with an
accuracy of $\sim 1$ ms.  
The present planning indicates
that the full detector will be installed by 2005.

From Nov. 1999 to June 2000 the Collaboration achieved a significant milestone in
the deployment and operation
of a ``demonstrator line'', instrumented with 7 PMTs at a depth of 1100~m
and connected to shore with a 37 km-long EOC. The shape of the atmospheric muon
zenith angular distribution was reproduced, despite the small number of PMTs
and their 1D-layout. The ANTARES relative and absolute
positioning acoustic system of rangemeters, compasses and tilt miters
was tested. Relative distances and absolute positioning 
were measured with an accuracy of $\sim 5$~cm and of $\sim 1$~m, respectively.

In Oct. 2001, the 40 km-long EOC for power and data transmission between 
the detector and the shore station in La Seyne sur Mer was deployed.
Since Dec. 9, 2002, the heart of the forthcoming array, the junction box 
(JB), has been in communication with the shore station. 
It was deployed during a sea operation requiring the dredging and lifting of 
2.5 km of the undersea EOC. 
On March 17, 2003 the first data were received from a prototype
detection line equipped with 15 OMs (5 storeys corresponding to 1/5 of an
ANTARES line). The line was deployed in Dec. 2002 and connected
in Mar. 2003 to the underwater JB using the Nautile 
manned submarine of the French IFREMER
oceanographic research agency. During the same mission, the Nautile
connected a prototype instrumentation line 
(deployed on Feb. 12, 2003)
incorporating instrumentation to monitor underwater environmental
parameters (a pulsed laser calibration system, 
a deep-sea Doppler current meter, detectors to measure sound velocity, 
salinity and water transparency).
The success of the submarine connections has proved the viability
of the final detector configuration with 12 line inter-connections radiating 
from the JB. Data are currently being acquired from PMTs, tilt meters and
compasses of the prototype line and from the instrumentation line. They
are consistent with the expected single counting rate of around 60 kHz 
due to $^{40}K$ $\beta$ decays and peaks in excess of 250 kHz 
from bioluminescence bursts.
The status and results of the prototype lines are described in~[4].
Fig.\,1 shows the layout of the future 12 line detector and
underwater photographs of the prototype lines. 

\section{Expected physics performances}

The ANTARES angular resolution is about 0.2$^{\circ}$ for 
$E_{\nu}> 10$~TeV, where pointing accuracy is not limited by the
$\nu-\mu$ kinematics, but by the PMT TTS and 
by light scattering in water. The ANTARES sensitivity (90\% c.l.)
for point-like source searches to the upgoing $\mu$ flux induced by a typical
$E^{-2}$ differential $\nu$ flux 
is in the range between $4 \div 50 \cdot 10^{-16}$ cm$^{-2}$ 
s$^{-1}$ after 1 yr. Search strategies and discovery potential are
discussed in~[5]. 
The energy resolution and methods to reconstruct muon energy and 
parent neutrino spectra are discussed in~[7]. The sensitivity 
(90\% c.l.) to $E^{-2}$ diffuse differential
neutrino fluxes achieved by rejecting the background with an energy cut 
of $E_{\mu} \ge 50$~GeV
is $8 \cdot 10^{-8}$ GeV cm$^{-2}$ s$^{-1}$ sr$^{-1}$.
The sensitivity of ANTARES in the search of WIMPs is given in~[8]. 
The upper limit on $\nu$-induced muon fluxes (90\% c.l.)  
from neutralino annihilation in the Sun 
is at the level of 400 km$^{-2}$ yr$^{-1}$ for $m_{\chi} \gtrsim 200$~GeV.
\section{References}
\vspace{-0.5cm}
\re
1.\ Amram P. \ et al.\ 2003, Astrop. Phys. {\bf 19}, 253
\re
2. Barwick S.W.\ et al., astro-ph/0211269
\re
3.\ Bouwhuis M.\ et al., 
{\it A data acquisition system for the ANTARES neutrino telescope}, 
this conference 
\re
4.\ Circella M.\ et al., 
{\it Toward the ANTARES neutrino telescope: results from a prototype line}, 
this conference
\re
5.\ Heijboer A.\ et al., 
{\it Point source searches with the ANTARES neutrino telescope},
this conference
\re
6.\ Hern\'andez-Rey J.\ et al., 
{\it Time Calibration of the ANTARES Neutrino Telescope}, this conference
\re
7.\ Romeyer \ et al., {\it Muon energy reconstruction in ANTARES and its
application to the diffuse neutrino flux}, 
this conference
\re
8.\ Thompson L.\ et al., 
{\it Dark matter searches with the ANTARES neutrino 
telescope},
this conference
\re
9.\ Waxman E., Bahcall J.N.\ 1999, Phys. Rev. {\bf D59} (1999) 023002
\re
10.\ Zornoza J.\ et al., 
{\it Study of photomultiplier tubes for the ANTARES neutrino telescope}, 
this conference
\begin{figure}[t]
  \begin{center}
    \includegraphics[height=6.5cm,width=11.5cm]{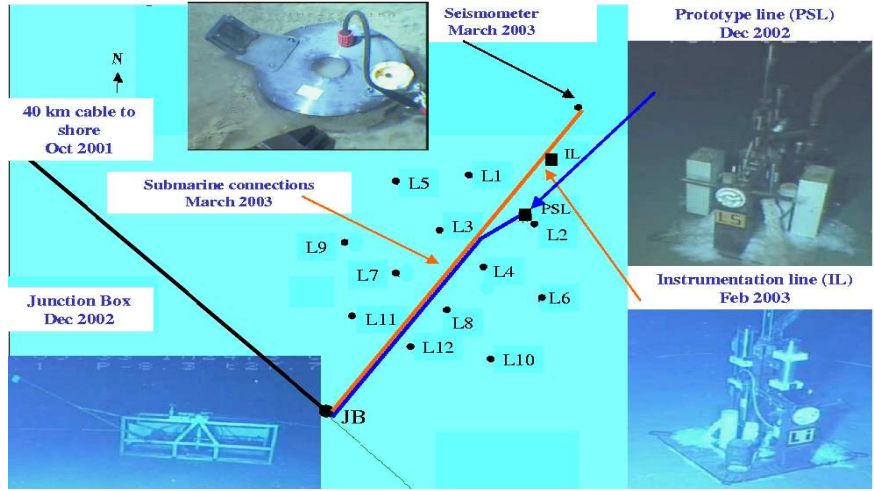}
  \end{center}
  \vspace{-1pc}
  \caption{The ANTARES layout: the positions of the 
deployed prototype lines (squares, PSL = Prototype Sector Line and IL =
Instrumentation Line),
the cables connecting them to the JB, the
40 km EOC and the future 12 lines are
indicated. 
Underwater photographs of the 2 lines, JB and seismometer 
are shown.}
\end{figure}
\endofpaper
\end{document}